# ON THE DYNAMICS OF INTERNAL WAVES PROPAGATING IN STRATIFIED MEDIA OF A VARIABLE DEPTH: EXACT AND ASYMPTOTIC SOLUTIONS

V.V.Bulatov, Yu.V.Vladimirov,

Institute for Problems in Mechanics RAS

Pr.Vernadskogo 101-1, 119526 Moscow, Russia

internalwave@mail.ru

**ABSTRACT**

The dynamics of internal waves in stratified media, such as the ocean or atmosphere, is highly dependent on the topography of their floor. A closed-form analytical solution can be derived only in cases when the water distribution density and the shape of the floor are modeled with specific functions. In a general case when the characteristics of stratified media and the boundary conditions are arbitrary, the dynamics of internal waves can be only approximated with numerical methods. However, numerical solutions do not describe the wave field qualitatively. At the same time, the need for a qualitative analysis of the far field of internal waves arises in studies applying remote sensing methods in space-based radar applications.. In this case, the dynamics of internal waves can be described using asymptotic models. In this paper, we derive asymptotic solutions to the problem of characterizing the far field of internal gravity waves propagating in a stratified medium with a smoothly varying floor.

**PROBLEM FORMULATION.**

In this study we consider a non-viscous incompressible nonhomogeneous medium. If it is unperturbed, we denote its density by $\rho(z)$ (the stratification is supposed to be stable, i.e. $\partial \rho / \partial z < 0$, the axis $z$ is directed downward from the medium surface). The system of the hydrodynamic equations takes the following form [1,2]

$$\frac{\partial \mathbf{v}}{\partial t} + (\mathbf{v} \cdot \nabla)\mathbf{v} = -\frac{\nabla p}{\rho} + \mathbf{g}, \; div\,\mathbf{v} = 0, \; \frac{\partial \rho}{\partial t} + \nabla \rho \cdot \mathbf{v} = 0 \qquad (1)$$

where $\mathbf{v} = (U_1, U_2, W)$ is the velocity vector, $\mathbf{g} = (0, 0, g)$ is the gravitation acceleration vector, $p$ and $\rho$ are the deviations of the pressure and the density from their equilibrium values. We consider a stratified medium with density $\rho_0(z)$, bounded in unperturbed state by the surface $z = 0$ and bottom $z = -H(y)$, with its depth depending solely on one horizontal variable $y$.

**EXACT AND ASYMPTOTIC SOLUTIONS.**

Then, restricting ourselves to the case of constant Vaisala-Brunt frequency $N^2(z) = -\frac{g}{\rho_0(z)} \frac{d\rho_0(z)}{dz} = const$, time law $\exp(-i\omega t)$ and the dependence of $x$ in the form of $\exp(ilx)$ in the Boussinesq approximation we shall obtain the following linearized about the quiescent state equation for the vertical components of velocity $W(z, y)$ (omitting the multiplier $\exp(ilx)$)

$$\frac{\partial^2 W}{\partial^2 z^2} - \frac{N^2 - \omega^2}{\omega^2} \frac{\partial^2 W}{\partial^2 y^2} + l^2 \frac{N^2 - \omega^2}{\omega^2} W = 0 \tag{2}$$

As the boundary conditions we take the "rigid lid" condition at $z = 0$ and no-fluid-loss condition at the bottom: $W = 0$ at $z = 0$, $W + \frac{dH(y)}{dy} U_2 = 0$ at $z = -H(y)$ where $(U_2, W)$ - velocity components. With relation to the function $H(y)$ we assume as follows: $H(y)$ is the continuously differentiable function with no more than a single minimum. The continuity variation $H(y)$ means that the relationship between the horizontal scale $L$ of variation $H(y)$ and the vertical scale $M$ is defined by the value $\lambda = L/M \gg 1$. In non-dimensional variables $x^* = x/L$, $y^* = y/L$; $z^* = z/M$, $h^*(y^*) = H(y)/M$; $l^* = lM$, $\omega^* = \omega/N$ the equation (2) and the boundary conditions are re-written as follows (the sign * is further-on omitted)

$$\frac{\partial^2 W}{\partial^2 z^2} - \frac{1}{\lambda^2 c^2} \frac{\partial^2 W}{\partial^2 y^2} + \frac{l^2}{c^2} W = 0 \tag{3}$$

$W = 0$, at $z = 0$ $\qquad W + \frac{dh(y)}{\lambda dy} U_2 = 0$, at $z = -h(y)$, $c^2 = \frac{\omega^2}{1 - \omega^2}$

Below we substitute the boundary condition at the bottom with large $\lambda$ by $W = 0$ at $z = -h(y)$. We shall seek the asymptotic solution of the problem (3.6.3) in the form typical for the geometric optics (WKBJ) method [3,4]

$$W = \left( F_0(z, y, \omega) + \frac{i}{\lambda} F_1(z, y, \omega) + \left(\frac{i}{\lambda}\right)^2 F_2(z, y, \omega) + ... \right) \exp(i\lambda S(y, \omega)) \tag{4}$$

where $F_m(z, y, \omega) = 0$ at $z = 0$ and at $z = -h(y)$, $m = 0,1,2,...$ Substituting the solution for $W$ in the form (4) into the equation (3) and making equal the members at $\lambda^0$ and $\lambda^1$, we obtain

$$\frac{\partial^2 F_0}{\partial^2 z^2} + \frac{S'^2 + l^2}{c^2} F_0 = 0, \quad F_0 = 0, \text{ at } z = 0 \text{ and at } z = -h(y) \tag{5}$$

$$\frac{\partial^2 F_1}{\partial^2 z^2} + \frac{S'^2 + l^2}{c^2} F_1 = \frac{1}{c^2}(2F_0'S' + F_0 S''), \quad F_1 = 0, \text{ at } z = 0 \text{ and at } z = -h(y)$$

where prime marks without indexes denote the derivatives over $y$. The solution of the first equation from (5) (the Sturm-Liouville problem) provides for the mode structure solution: the dispersion relationships $\kappa_n^2(y, \omega) = \frac{c^2 n^2 \pi^2}{h^2(y)}$, $n = 1,2,...$ and the eigenfunctions in a zero-order approximation (vertical modes) $F_{0n}(z, y, \omega) = A_{0n}(y, \omega) \sin(n\pi z / h(y))$, $n = 1,2,...$ . The eikonal $S_n(y, \omega)$ is defined from the relationship $\kappa_n^2(y, \omega) = S_n'^2(y, \omega) + l^2$. To find the amplitude $A_{0n}(y, \omega)$ we use the resolvability condition for the second equation from (5), which requires the orthogonality of the equation's right member and of the function $F_{0n}$. By multiplying at fixed $n$ this equation by $F_{0n}$ and integrating over $z$ from 0 to $h(y)$, we obtain the following «conservation law» $\frac{\partial}{\partial y}\left(\int_0^{h(y)} F_{0n}^2(z, y, \omega)dz \cdot S'(y, \omega)\right) = 0$. By integrating the eigenfunction and considering the equation for the eikonal we finally get: $A_{0n} = \frac{B_{0n}(y_0, \omega)}{\sqrt[4]{c^2 n^2 \pi^2 - h^2(y)l^2}}$, where the variable $B_{0n}$ depends on $\omega$ and the initial eikonal value at any point $y_0$, $S_n(y_0, \omega)$. The eikonal $S_n(y, \omega)$ shall be defined as $S_n(y, \omega) = \int_y^{y^*} \sqrt{\kappa_n^2(y, \omega) - l^2} dy$, where $y^*$ is the «turning point», i.e. the root of equation $\kappa_n^2(y, \omega) = l^2$. Then WKBJ solution for a separate wave mode is given as follows

$$W_n^{\pm}(z, y) = \frac{D_{\pm}}{\sqrt[4]{c^2 n^2 \pi^2 - h^2(y)l^2}} \exp(\pm i\lambda(S_n(y, \omega))) \sin\frac{n\pi z}{h(y)}$$

where plus-sign corresponds to an «incident» wave, and the minus-sign corresponds to a reflected wave. A geometric solution is not working near the turning point (the amplitude $A_{0n}$ tends to infinity). The solution's uniform asymptotics for a individual wave mode applicable at the turning point is expressed by the Airy function and is given in the form

$$W_n = \frac{\sqrt{2}\pi \left(\frac{3}{2}\lambda S_n(y,\omega)\right)^{1/6}}{\lambda^{1/2} \sqrt[4]{c^2 n^2 \pi^2 - h^2(y)l^2}} Ai\left(\left(\frac{3}{2}\lambda S_n(y,\omega)\right)^{2/3}\right) \sin\frac{n\pi z}{h(y)}$$

where $Ai(x)$ is the Airy function. The complete solution in the WKBJ approximation for a single mode before the turning point (i.e. within the «wave region») appears to be

$$W_n = \frac{\sqrt{2\pi}}{\lambda^{1/2} \sqrt[4]{c^2 n^2 \pi^2 - h^2(y)l^2}} \cos\left(\lambda S_n(y,\omega) - \frac{\pi}{4}\right) \sin\frac{n\pi z}{h(y)}$$

and beyond the turning point (in the region of exponential fading)

$$W_n = \frac{\sqrt{\pi}}{\lambda^{1/2} \sqrt[4]{c^2 n^2 \pi^2 - h^2(y)l^2}} \exp(-\lambda|S_n(y,\omega)|) \sin\frac{n\pi z}{h(y)}$$

For a linear bottom profile this problem we solved analytically. The solution for a individual wave mode is given by the Macdonald function $K_\nu$ of imaginary index as follows

$$W_n = e^{-i\frac{\pi\nu}{2}} K_\nu\left(l\sqrt{\lambda^2 y^2 - \frac{z^2}{c^2}}\right) \sin\left(\frac{n\pi}{\ln\Delta} \ln\frac{\lambda c y - z}{\lambda c y + z}\right)$$

where $\Delta = \frac{\lambda c + 1}{\lambda c - 1}$, $\nu = \frac{2\pi n i}{\ln\Delta}$, $c^2 = \frac{\omega}{1-\omega^2}$, $\lambda = \frac{1}{\gamma}$ $\gamma$ is the bottom's inclination, $h(y) = -y$.

### NUMERICAL RESULTS AND DISCUSSIONS.

The figures show the results of internal gravity wave vertical velocity calculations for two non-linear stratified medium bottom shape. Numerical calculations show that the two types of existent waves: captured waves and progressive waves. In fig.1 results of a function $W_1(y,z)$ calculations are presented. In fig.2 results of a function $W_2(y,z)$ calculations are presented. The stratified medium bottom shapes are shown in figures. The effect of spatial-frequency "blocking" of the wave field was revealed. Depending on the wave and bottom shape the far internal waves field can be localized in a bounded region of space (captured waves), or spread, in the absence of turning points, over a large distance compared to the medium depth (progressive waves). The region, which progressive wave can penetrate, is completely determined by the presence of turning points, whose locations depend on the medium stratification and bottom topography. In this paper asymptotic and exact representations of solutions was obtained that describe the far field of internal gravity waves in a stratified mediums of variable depth. Using developed asymptotic methods, one can consider a wide class of interesting physical problems, including problems concerning the propagation of internal gravity wave packets in non-homogeneous stratified media under the assumption that the modification of the parameters of a vertically stratified medium are slow in the horizontal direction. Numerical analyses that are performed using typical ocean parameters reveal that actual dynamics of the internal gravity waves are strongly influenced by horizontal non-homogeneity of the ocean bottom. In this paper we use an analytical approach, which avoids the numerical calculation widely used in analysis of internal gravity wave dynamics in stratified mediums. Asymptotic solutions, which are obtained in this paper, can be of significant importance for engineering applications, since the method of geometrical optics, which we modified in order to calculate the wave field near caustic, makes it possible to describe different wave fields in a rather wide class of other problems.

The asymptotic representations constructed in this paper allow one to describe the far field of the internal gravity waves generated by a source moving over a slowly varying bottom. The obtained asymptotic expressions for the solution are uniform and reproduce fairly well the essential features of wave fields near caustic surfaces and wave fronts. In this paper the problem of reconstructing non-harmonic wave packets of internal gravity waves generated by a source moving in a horizontally stratified medium is considered. The solution is proposed in terms of modes, propagating independently in the adiabatic approximation, and described as a non-integer power series of a small parameter characterizing the stratified medium. In this study we analyze the evolution of non-harmonic wave packets of internal gravity waves generated by a moving source under the assumption that the parameters of a vertically stratified medium (e.g. an ocean) vary slowly in the horizontal direction, as compared to the characteristic length of the density. A specific form of the wave packets, which can be parameterized in terms of model functions, e.g. Airy functions, depends on local behavior of the dispersion curves of individual modes in the vicinity of the corresponding critical points.

In this paper a modified space-time ray method is proposed, which belongs to the class of geometrical optics methods [1-3]. The key point of the proposed technique is the possibility to derive the asymptotic representation of the solution in terms of a non-integer power series of the small parameter $\varepsilon = \lambda/L$, where $\lambda$ is the characteristic wave length, and L is the characteristic scale of the horizontal heterogeneity. The explicit form of the asymptotic solution was determined based on the principles of locality and asymptotic behavior of the solution in the case of a stationary and horizontally homogeneous medium. The wave packet amplitudes are determined from the energy conservation laws along the characteristic curves. A typical assumption made in studies on the internal wave evolution in stratified media is that the wave packets are locally harmonic. A modification of the geometrical optics method, based on an expansion of the solution in model functions, allows one to describe the wave field structure both far from and at the vicinity of the wave front.

Using the asymptotic representation of the wave field at a large distance from a source moving in a layer of constant depth, we solve the problem of constructing the uniform asymptotics of the internal waves in a medium of varying depth. The solution is obtained by modifying the previously proposed "vertical modes-horizontal rays" method, which avoids the assumption that the medium parameters vary slowly in the vertical direction. The solution is parameterized through the Airy waves. This allows one to describe not only the evolution of the non-harmonic wave packets propagating over a slow-varying fluid bottom, but also specify the wave field structure associated with an individual mode both far from and close to the wave front of the mode. The Airy function argument is determined by solving the corresponding eikonal equations and finding vertical spectra of the internal gravity waves. The wave field amplitude is determined using the energy conservation law, or another adiabatic invariant, characterizing wave propagation along the characteristic curves.

Modeling typical shapes and stratification of the ocean shelf, we obtain analytic expressions describing the characteristic curves and examine characteristic properties of the wave field phase structure. As a result it is possible to observe some peculiarities in the wave field structure, depending on the shape of ocean bottom, water stratification and the trajectory of a moving source. In particular, we analyze a spatial blocking effect of the low-frequency components of the wave field, generated by a source moving alongshore with a supercritical velocity. Numerical analyses that are performed using typical ocean parameters reveal that actual dynamics of the internal gravity waves are strongly influenced by horizontal nonhomogeneity of the ocean bottom. In this paper we use an analytical approach, which avoids the numerical calculation widely used in analysis of internal gravity wave dynamics in stratified ocean.

In this paper the most difficult question is considered that can appear when we investigate the problems of wave theory with the help of geometrical optics methods and its modifications. And the main question consists in finding of asymptotic solution near special curve (or surface), which is called caustic. It is well known, that caustic is an envelope of a fam-ily of rays, and asymptotic solution is obtained along these rays. Asymptotic representation of the field describe qualitative change of the wave field, and that is description of the field, when we cross the area of "light", where wave field exists, and come in the area of "shadow", where we consider wave field to be rather small. Each point of the caustic corresponds to a specified ray, and that ray is tangent at this point.

It is a general rule that caustic of a family of rays single out an area in space, so that rays of that family cannot appear in the marked area. There is also another area, and each point of that area has two rays that pass through this point. One of those rays has already passed this point, and another is going to pass the point. Formal approximation of geometrical optics or WKB approximation cannot be applied near the caustic, that is because rays merge together in that area, after they were reflected by caustic. If we want to find wave field near the caustic, then it is necessary to use special approximation of the solution, and in the paper a modified ray method is proposed in order to build uniform asymptotic expansion of integral forms of the internal gravity wave field. After the rays are reflected by the caustic, there appears a phase shift. It is clear that the phase shift can only happen in the area where methods of geometrical optics, which were used in previous sections, can't be applied. If the rays touch the caustic several times, then additional phase shifts will be added. Phase shift, which was created by the caustic, is rather small in comparison with the change in phase along the ray, but this shift can considerably affect interference pattern of the wave field.

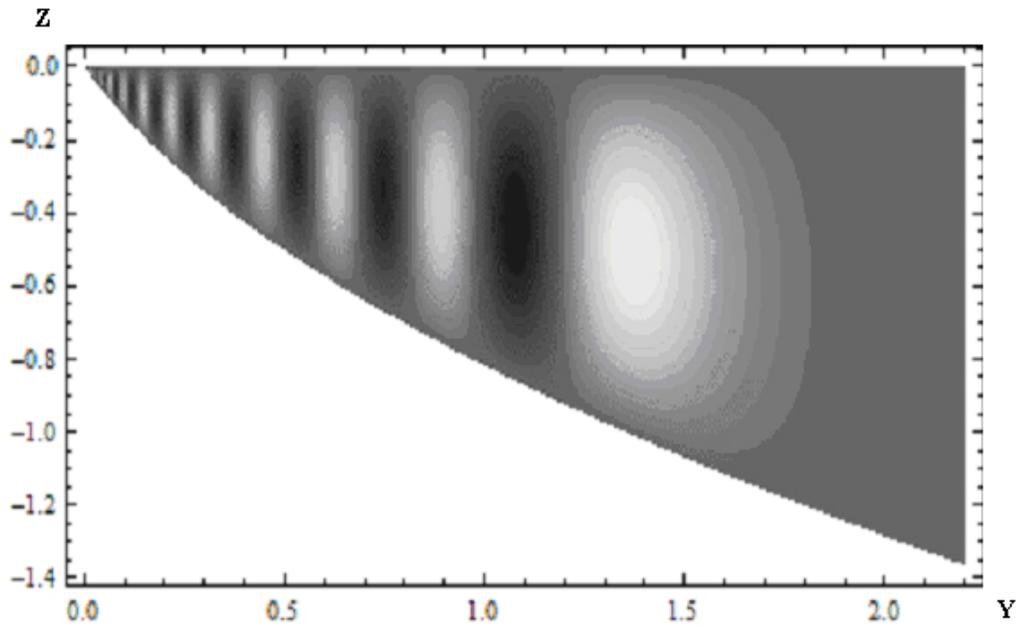

*Fig.1 Internal gravity waves first mode for infinitely descending bottom profile (captured waves)*

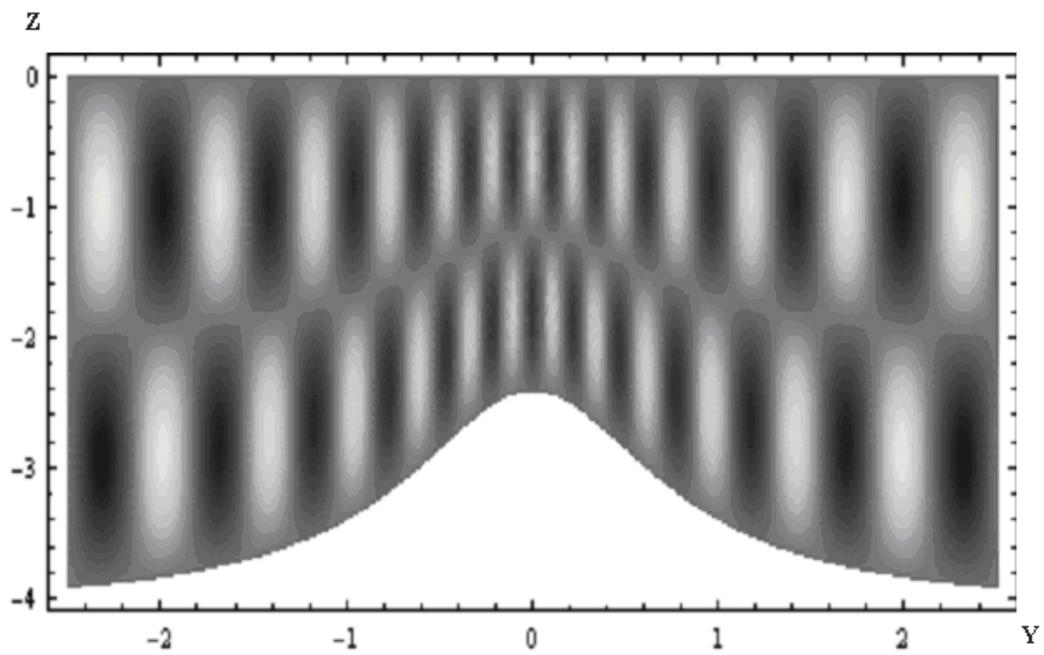

*Fig.2 Internal gravity waves second mode for hill bottom profile (progressive waves).*

ACKNOWLEDGMENTS

The results presented in the paper have been obtained by research performed under projects supported by the Russian Foundation for Basic Research (No.11-01-00335a "Stability and Instability of Flows with Interfaces", No. 12-05-00227a "Dynamics of Wave Processes in Conditions of the Arctic Basin"), Program of the Russian Academy of Sciences "Fundamental Problems of Oceanology: Physics, Geology, Biology, Ecology" .

APPLICATIONS

The results of this paper represent significant interest for physics and mathematics. Besides, asymptotic solutions, which are obtained in this paper, can be of significant importance for engineering applications, since the method of geometrical optics, which we modified in order to calculate the wave field near caustic, makes it possible to describe different wave fields in a rather wide class of other problems.